\documentclass[conference]{IEEEtran}
\IEEEoverridecommandlockouts
\usepackage{amsmath,amssymb,amsfonts}
\usepackage{algorithmic}
\usepackage{graphicx}
\usepackage{textcomp}
\usepackage{xcolor}
\usepackage{hyperref}
\usepackage{makecell}
\usepackage{algorithm}
\usepackage{algorithmic}
\usepackage{booktabs}
\usepackage{cite}
\usepackage{array,tabularx}
\usepackage{threeparttable}
\newcolumntype{C}{>{\centering\arraybackslash}X}  % centred, stretchable column
\hypersetup{draft}

\usepackage[top=0.74in, bottom=1in, left=0.75in, right=0.75in]{geometry}

\hypersetup{
    colorlinks=true, % Use colored links instead of boxes
    linkcolor=black,  % Color of internal links
    citecolor=black, % Color of citation links
    filecolor=black, % Color of file links
    urlcolor=black    % Color of external links
}

\def\BibTeX{{\rm B\kern-.05em{\sc i\kern-.025em b}\kern-.08em
    T\kern-.1667em\lower.7ex\hbox{E}\kern-.125emX}}
\begin{document}

\title{Quantum Machine Learning for UAV Swarm Intrusion Detection\thanks{The views expressed in this article are those of the authors and do not represent the views of Wells Fargo. This article is for informational purposes only. Nothing contained in this article should be construed as investment advice. Wells Fargo makes no express or implied warranties and expressly disclaims all legal, tax, and accounting implications related to this article. }}

\author{
\IEEEauthorblockN{
    Kuan-Cheng Chen\IEEEauthorrefmark{2}\IEEEauthorrefmark{3}\IEEEauthorrefmark{1},
    Samuel Yen-Chi Chen \IEEEauthorrefmark{4},
    Tai-Yue Li \IEEEauthorrefmark{5},
    Chen-Yu Liu \IEEEauthorrefmark{6},
    Kin K. Leung\IEEEauthorrefmark{2}
}
\IEEEauthorblockA{\IEEEauthorrefmark{2}Department of Electrical and Electronic Engineering, Imperial College London, London, UK}
\IEEEauthorblockA{\IEEEauthorrefmark{3}Centre for Quantum Engineering, Science and Technology (QuEST), Imperial College London, London, UK}
\IEEEauthorblockA{\IEEEauthorrefmark{4}Wells Fargo, New York, USA}
\IEEEauthorblockA{\IEEEauthorrefmark{5}National Center for High-performance Computing, Hsinchu, Taiwan}
\IEEEauthorblockA{\IEEEauthorrefmark{6}Graduate Institute of Applied Physics, National Taiwan University, Taipei, Taiwan}

\thanks{*Corresponding Author: kuan-cheng.chen17@imperial.ac.uk}
}

\maketitle

\begin{abstract}
Intrusion detection in unmanned-aerial-vehicle (UAV) swarms is complicated by high mobility, non-stationary traffic, and severe class imbalance. Leveraging a 120 k-flow simulation corpus that covers five attack types, we benchmark three quantum-machine-learning (QML) approaches—quantum kernels, variational quantum neural networks (QNNs), and hybrid quantum-trained neural networks (QT-NNs)—against strong classical baselines. All models consume an 8-feature flow representation and are evaluated under identical preprocessing, balancing, and noise-model assumptions. We analyse the influence of encoding strategy, circuit depth, qubit count, and shot noise, reporting accuracy, macro-F1, ROC-AUC, Matthews correlation, and quantum-resource footprints. Results reveal clear trade-offs: quantum kernels and QT-NNs excel in low-data, nonlinear regimes, while deeper QNNs suffer from trainability issues, and CNNs dominate when abundant data offset their larger parameter count. The complete codebase and dataset partitions are publicly released to enable reproducible QML research in network security.
\end{abstract}

\begin{IEEEkeywords}
Quantum Machine Learning,  Quantum Kernels, Quantum Neural Networks, Intrusion Detection Systems, Unmanned Aerial Vehicle Networks, IoTs
\end{IEEEkeywords}

\section{Introduction}
\label{sec:intro}
  Unmanned aerial vehicle (UAV) swarms coordinate collections of airborne nodes through distributed control and shared wireless links to achieve wide-area sensing, coverage, and mission resilience\cite{campion2018uav}. Compared with single-UAV systems, swarms benefit from redundancy, cooperative perception, and scalable area coverage, but they also face tighter constraints on energy, link reliability, and real-time decision-making\cite{zeng2018wireless,chen2020review}. Architecturally, swarms range from centralized control to fully distributed formations, with hybrid designs common in practice\cite{bertizzolo2020swarmcontrol}. Communications leverage cellular (4G/5G)\cite{zeng2018wireless}, satellite\cite{fotouhi2019survey}, and WLANs\cite{menouar2017uav}; routing spans topology-based\cite{bekmezci2013flying}, location-aware\cite{asadpour2014micro}, and swarm-intelligence protocols\cite{motlagh2016low}; and navigation integrates localization, path planning, collision avoidance, and formation control. This operational stack yields agility and robustness, but it also broadens the attack surface in contested or resource-limited environments.

The security posture of UAV swarms is shaped by open wireless media, dynamic multi-hop routing, limited computation/power, and intermittent connectivity. Consequently, attacks span (i) communication-layer threats such as eavesdropping and jamming\cite{choudhary2018intrusion}; (ii) identity threats including impersonation, replay, man-in-the-middle, and Sybil attacks\cite{choudhary2018intrusion}; (iii) resource exhaustion via DoS/DDoS, malware, or hijacking\cite{choudhary2018intrusion}; (iv) routing manipulation such as wormhole, blackhole, and gray-hole behaviors\cite{wang2024survey}; (v) data integrity violations, including tampering and false-data injection (e.g., GPS spoofing)\cite{wang2024survey}; and (vi) machine learning (ML)-centric threats targeting models used for perception or control (poisoning, backdoors, and adversarial examples\cite{lopez2021towards}. Survey evidence indicates that many proposed defenses presume partial channel knowledge, static adversaries, or generous resource budgets that rarely hold in fast-moving swarms\cite{choudhary2018intrusion,lopez2021towards}. These constraints underscore the need for lightweight, data-driven intrusion detection capable of operating under mobility, class imbalance, and non-stationary traffic.

Intrusion Detection Systems (IDS) complement cryptography and physical-layer hardening by monitoring network flows and protocol behaviors to surface anomalies or signatures in near-real time\cite{choudhary2018intrusion}. In swarms, IDS must respect limited energy and compute budgets, cope with rapidly changing topologies, and remain robust to imbalanced attack distributions. Flow-level learning is particularly attractive because it avoids payload inspection, is protocol-agnostic, and can be federated or decentralized when backhaul is constrained\cite{santos2023flow}. However, classical models may require extensive feature engineering or large labeled datasets to capture complex, nonlinear boundaries among attack types\cite{mrad2021federated}.

\begin{figure*}[!t]
\centering
\includegraphics[width=0.7\linewidth]{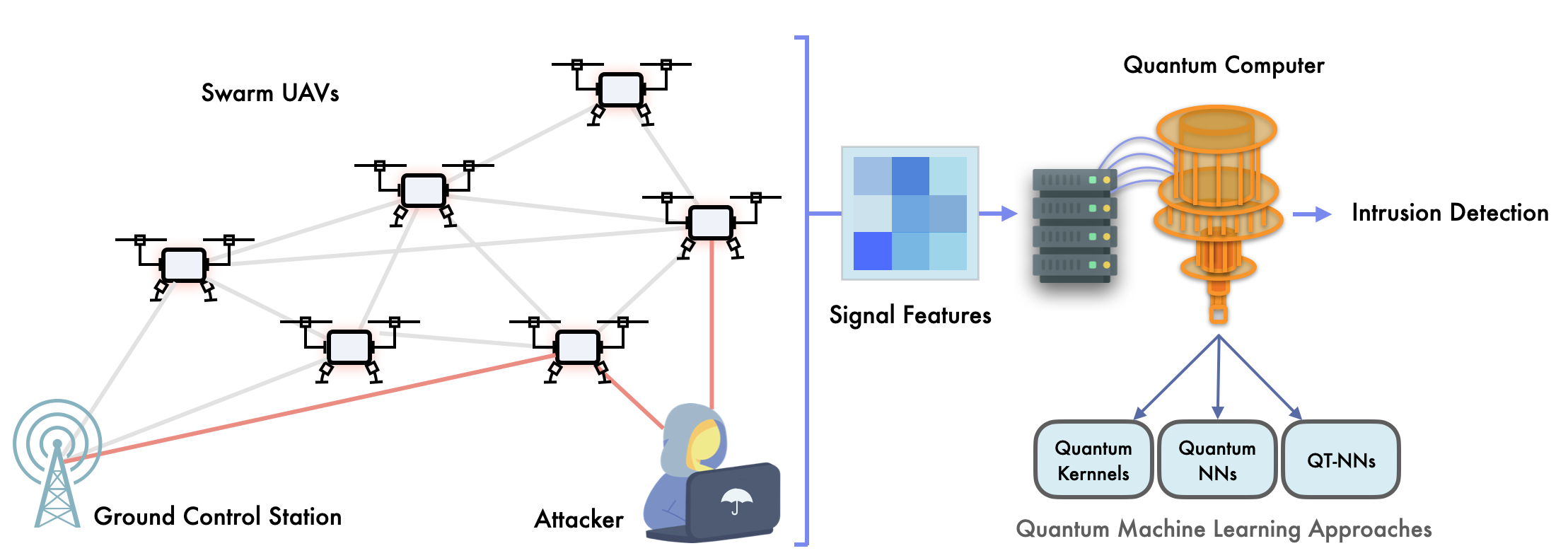}
\caption{
Overview of the QML for UAV swarm intrusion detection. 
}
 \label{fig:scheme}
\end{figure*}

Recent advances in quantum computing show promise for optimization\cite{abbas2024challenges}, simulation\cite{daley2022practical}, and high-dimensional learning\cite{cerezo2022challenges}--domains long bounded by classical resource limits. Early results in molecular simulation\cite{chen2025exploring}, combinatorial optimization\cite{chen2025resource}, and Quantum Machine Learning (QML)\cite{chen2024quantum,chen2020variational,chen2022quantum,chen2021federated,chen2025compressedmediq,chen2025consensus,chen2024quantum2}, which leverages quantum circuits to encode and process data in high-dimensional Hilbert spaces. By exploiting superposition, entanglement, and interference, QML aspires to improve expressive capacity, representation power, and scaling\cite{chen2025validating,liu2024towards}. In security--critical, resource-constrained settings such as intrusion detection for UAV swarm networks--where data can be non-stationary, imbalanced, and noisy--QML may enhance generalization and robustness\cite{ma2025robust,chen2025noise}. This work therefore examines whether hybrid quantum–classical models--specifically quantum kernel methods\cite{hsu2025quantum}, quantum neural networks (QNNs)\cite{abbas2021power,yu2024shedding}, and quantum-trained neural networks (QT-NNs)\cite{liu2024quantum,liu2025quantum,liu2025programming,chen2025quantum}--can deliver competitive or superior intrusion-detection performance on a realistic UAV swarm security dataset.

In this article, we benchmark quantum kernels, QNNs, and QT-NNs on the UAVIDS-2025 dataset\cite{zenguavids2025}, a comprehensive testbed for intrusion detection in UAV swarm networks. After domain-relevant feature engineering to obtain an interpretable 8-dimensional representation, we construct a balanced subset and systematically evaluate each quantum approach under realistic data conditions. We compare accuracy, robustness, and resource costs against classical baselines--support vector machines (SVMs). Our findings delineate where QML offers tangible advantages, where classical models remain competitive, and which quantum design choices most strongly influence performance in UAV-based security scenarios.

\section{PRELIMINARIES}

\subsection{Quantum States and Operators}

\subsubsection{Hilbert Space and Qubits}

A quantum system of \(N\) qubits is described by a complex Hilbert space \(\mathcal{H} \cong \mathbb{C}^{2^N}\). Any pure state \(|\psi\rangle\) in \(\mathcal{H}\) is a normalized vector, \(\|\psi\| = 1\). However, due to decoherence and interactions with the environment, a general description of the system may require the concept of a density operator \(\rho\). Formally, \(\rho\) is a positive semidefinite, Hermitian operator of trace one:
\begin{equation}
\rho \succeq 0, \quad \rho = \rho^\dagger, \quad \mathrm{Tr}(\rho) = 1.
\end{equation}
A density operator with \(\mathrm{rank}(\rho)=1\) corresponds to a pure state \(\rho = |\psi\rangle\langle \psi|\). States with higher rank are known as mixed states and can be interpreted as probabilistic ensembles of pure states.

\subsubsection*{Measurements and Observables}

Observables in quantum mechanics are represented by Hermitian operators \(O\) acting on \(\mathcal{H}\). Measuring \(\rho\) with respect to \(O\) yields an expectation value \(\mathrm{Tr}(O\,\rho)\). When making binary decisions (e.g., labeling data as \(+1\) or \(-1\)), one often uses projective or POVM (Positive Operator-Valued Measure) measurements. In learning contexts, one might designate a measurement operator \(O\) to map a quantum state to a real-valued label:
\begin{equation}
y := \mathrm{Tr}\bigl(O\,\rho\bigr), \quad \|O\|_2 \le 1,
\end{equation}
where \(\|\cdot\|_2\) denotes the spectral norm, bounding the maximal eigenvalue of \(O\).

\subsection{Quantum Kernel Methods}

Classical kernel methods (e.g., support vector machines, kernel ridge regression) rely on feature maps \(\phi: X \to F\) from an input space \(X\) into a high-dimensional feature space \(F\). A quantum kernel method adopts a quantum feature map, using an \(N\)-qubit circuit that embeds data into \(\mathcal{H}\) and leverages quantum measurement to compute inner products.

Let \(x \in X \subset \mathbb{R}^d\) be a classical data vector. A parameterized unitary operator
\begin{equation}
U_E(\mathbf{x}): |0\rangle^{\otimes N} \mapsto |\phi(\mathbf{x})\rangle \in \mathcal{H}
\end{equation}
encodes \(x\) into a quantum state. The resulting density operator is
\begin{equation}
\rho(\mathbf{x}) = |\phi(\mathbf{x})\rangle\langle \phi(\mathbf{x})|.
\end{equation}
Because \(|\phi(\mathbf{x})\rangle\) may occupy a large Hilbert space (dimension \(2^N\)), quantum feature maps are hypothesized to capture complex relationships among data that are difficult to replicate classically.

The quantum kernel function between two samples \(x\) and \(x'\) is defined as
\begin{equation}
K(\mathbf{x}, \mathbf{x}') = \mathrm{Tr}\bigl[\rho(\mathbf{x})\,\rho(\mathbf{x}')\bigr] = \bigl|\langle \phi(\mathbf{x}) \mid \phi(\mathbf{x}')\rangle\bigr|^2.
\label{eq:qkernel_def}
\end{equation}
This overlap can be experimentally estimated by building an interference circuit that effectively compares \(|\phi(\mathbf{x})\rangle\) and \(|\phi(\mathbf{x}')\rangle\). For each pair \(x, x'\) in the training set, one obtains an empirical estimate \(\widehat{K}(x,x')\). Collecting these values into a matrix \(K\), we arrive at an \(n \times n\) Gram matrix of quantum-induced inner products.

Once \(K\) is available, classical machine learning techniques (SVM, Gaussian processes, kernel ridge regression, etc.) handle the final optimization. For instance, in a regularized regression approach with labels \(\{y_i\}\),
\begin{equation}
\omega^* = \underset{\omega}{\mathrm{argmin}} \, \sum_{i=1}^n \Bigl( y_i - \mathrm{Tr}\bigl[\rho(x_i)\,\omega\bigr] \Bigr)^2 + \lambda\,\|\omega\|^2,
\end{equation}
where \(\omega\) can be represented in the span of \(\{\rho(x_i)\}\). Equivalently, in the dual formulation, \(\omega^*\) is related to \((K + \lambda I)^{-1}\mathbf{y}\). Thus, the model prediction for a new \(x\) is given by
\begin{equation}
h(x) = \mathrm{sign}\Bigl[\mathrm{Tr}\bigl(\rho(x)\,\omega^*\bigr)\Bigr] = \mathrm{sign}\Bigl[K(x,\cdot)\,(K + \lambda I)^{-1}\,\mathbf{y}\Bigr],
\end{equation}
up to saturation bounds (e.g., \(\pm 1\) for classification).

\paragraph*{Encoding choices and estimation.}
In practice, \(U_E(\mathbf{x})\) may use \emph{angle encoding} (e.g., rotations \(R_\alpha(\kappa x_j)\)) or \emph{amplitude encoding} (loading a normalized vector into amplitudes). Kernel entries are estimated from \(M\) circuit shots with standard deviation \(O(1/\sqrt{M})\). Finite-shot noise and readout errors bias the Gram matrix; regularization (diagonal jitter), centering, and error mitigation help maintain positive semidefiniteness and numerical stability.

\subsection{Quantum Neural Network (QNN)}

A Quantum Neural Network (QNN) is a parameterized quantum model that maps classical inputs to measurement statistics of a quantum circuit. From a physics perspective, it prepares a family of quantum states whose expectation values serve as sufficient statistics for inference; from a computer science perspective, it is a differentiable hypothesis class with tunable parameters and well-defined training objectives.

\subsubsection*{Model}

Let \(U_E(\mathbf{x})\) be a data-encoding (feature) unitary and \(U(\boldsymbol{\theta})\) a trainable, layered unitary (``quantum layer''), possibly repeated with data re-uploading. The QNN is defined as,

\begin{align}
|\psi(\mathbf{x};\boldsymbol{\theta})\rangle 
&= U(\boldsymbol{\theta})\, U_E(\mathbf{x})\, |0\rangle^{\otimes N}, \label{eq:qnn_state} \\
\mathbf{f}(\mathbf{x};\boldsymbol{\theta}) &= \bigl(\langle O_1\rangle, \ldots, \langle O_C\rangle \bigr),\\ 
& \langle O_c\rangle \equiv \langle \psi(\mathbf{x};\boldsymbol{\theta})|\, O_c \,|\psi(\mathbf{x};\boldsymbol{\theta})\rangle. \label{eq:qnn_scores}
\end{align}

Here, \(\{O_c\}_{c=1}^C\) are measurement observables (e.g., local Pauli operators or short-depth Pauli sums) that act as \emph{readout heads}. For binary classification one may use a single head with thresholding; for multi-class tasks, \(\mathbf{f}\) is mapped to probabilities via a classical link function (e.g., softmax). A hybrid post-processing layer is optional: \(\hat{\mathbf{y}} = \sigma(W\,\mathbf{f} + b)\), with small \(W, b\), which preserves the quantum feature map while enabling calibrated outputs.

\paragraph{Encoding and expressivity.}

Common encodings include angle encodings (\(R_Y, R_Z\) rotations), basis encodings, and amplitude-style embeddings (with higher state-preparation cost). Expressivity is controlled by (i) the entangling pattern and depth of \(U(\boldsymbol{\theta})\), (ii) data re-uploading frequency, and (iii) the locality of \(\{O_c\}\). These choices trade off representation power against trainability on noisy, shallow hardware.

\subsubsection*{Training Objectives}

Given labels \(y_i\) and predictions \(\hat{\mathbf{p}}_i = \mathrm{softmax}(\mathbf{f}(\mathbf{x}_i;\boldsymbol{\theta}))\), a standard objective is cross-entropy with weight decay:
\begin{equation}
\mathcal{L}(\boldsymbol{\theta}) = -\frac{1}{n}\sum_{i=1}^n \log \hat{p}_{i,y_i} + \frac{\lambda}{2}\|\boldsymbol{\theta}\|_2^2.
\end{equation}

For binary tasks, one may use logistic loss on a single head \(f(\mathbf{x};\boldsymbol{\theta})\). Margin-based losses (e.g., hinge) or robust alternatives can be adopted when class imbalance or label noise is significant.

\subsubsection*{Gradients and Optimization}

For gates generated by Pauli operators, the parameter-shift rule gives unbiased gradients using two shifted circuit evaluations per parameter:
\begin{equation}
\frac{\partial \langle O_c \rangle}{\partial \theta_k} = \frac{1}{2} \left[ \langle O_c \rangle_{\theta_k + \frac{\pi}{2}} - \langle O_c \rangle_{\theta_k - \frac{\pi}{2}} \right].
\end{equation}

Gradients propagate through the classical post-processing exactly (backprop-through-expectations). Stochastic optimizers (e.g., Adam) with mini-batches are typically used; finite-shot readout introduces gradient noise proportional to \(1/M\), where \(M\) is the number of shots per expectation.

\subsubsection*{Ansatz Design and Trainability}

Two complementary principles guide \(U(\boldsymbol{\theta})\):
\begin{itemize}
    \item \emph{Hardware-efficient layers}: Interleaved single-qubit rotations and native entanglers arranged to match device connectivity; parameter count scales as \(\mathcal{O}(NL)\) with qubit count \(N\) and depth \(L\).
    \item \emph{Problem-informed layers}: Symmetry-preserving or locality-aware motifs aligned with the structure of the input (e.g., feature invariances), which can improve sample efficiency and reduce overfitting.
\end{itemize}

Deep random circuits with global cost functions may exhibit \emph{barren plateaus} (exponentially vanishing gradient variance with \(N\)). Practical mitigations include local cost functions, symmetry-preserving initializations, layer-wise training, entanglement sparseness, and data re-uploading to increase expressivity without excessive depth.

\paragraph{Interpretation.}

The QNN implements a learned quantum feature map \(\Phi_{\boldsymbol{\theta}}: \mathbf{x} \mapsto \rho(\mathbf{x};\boldsymbol{\theta})\) followed by linear functionals \(\mathrm{Tr}[\rho(\mathbf{x};\boldsymbol{\theta})\, O_c]\). This realizes a nonclassical representation whose geometry—induced by the Fubini–Study metric and circuit symmetries—can support expressive and generalizable decision boundaries under realistic circuit constraints.

\subsection{Quantum-Trained Neural Networks}

Quantum-Trained Neural Networks (QT-NN) integrate a parameterized quantum circuit as a \emph{weight generator} or \emph{adapter} for a classical neural network. From a physics perspective, the quantum module prepares an entangled state whose measurement statistics induce structured priors and stochasticity over the classical model’s parameters. From a computer-science perspective, QT-NN is a hybrid hypernetwork: a differentiable quantum map produces (parts of) the weights of a classical network that is trained end-to-end on the task loss.

\subsubsection*{Model}

QT-NN leverages a quantum circuit \(U(\boldsymbol{\theta})\) to prepare quantum states directly from the trainable angles $\boldsymbol{\theta}$ without the requirement to encode the classical input $\mathbf{x}$. The resulting measurement vector $\mathbf{g}(\boldsymbol{\theta})$, which includes the probabilities of each computational basis, is used not for direct classification, but to generate classical parameters via a structured mapping \(\mathcal{T}\), such as normalization, scaling, or low-rank transformation:
\begin{equation}
\mathbf{w} = \mathcal{T}\big(\mathbf{g}(\boldsymbol{\theta})\big)\ \cup\ \mathbf{w}_{\text{base}}.
\end{equation}
These parameters define a classical network \(f_{\mathbf{w}}: \mathbb{R}^d \to \mathbb{R}^C\), yielding predictions \(\hat{\mathbf{y}} = f_{\mathbf{w}}(\boldsymbol{x})\). QT-NN may produce full weight layers, lightweight adapters (e.g., rank-1 updates), or modulate existing weights, with quantum entanglement providing an inductive bias over generated parameters.

\subsubsection*{Training Objective and Optimization}
QT-NN is trained end-to-end on a task loss \(\mathcal{L}(\boldsymbol{\theta},\mathbf{w}_{\text{base}})\) (e.g., cross-entropy). Gradients propagate through the classical network by standard backpropagation and through the quantum generator via parameter-shift:
\begin{equation}
\frac{\partial \langle O_k\rangle}{\partial \theta_j}=\tfrac{1}{2}\!\left[\langle O_k\rangle_{\theta_j+\frac{\pi}{2}}-\langle O_k\rangle_{\theta_j-\frac{\pi}{2}}\right].
\end{equation}
Finite-shot sampling induces controlled stochasticity on the generated weights, which can aid robustness under class imbalance and distribution shift. When \(\mathbf{g}\) is input-dependent, QT-NN realizes a context-conditioned quantum hypernetwork; when input-independent, it functions as a learned quantum prior over parameters.

\subsubsection*{Resource, Complexity, and Design Considerations}
Let \(K\) be the number of measured observables and \(L\) the circuit depth. A forward pass costs \(\mathcal{O}(K L)\) circuit executions per shot; grouping commuting observables reduces overhead. Crucially, when the quantum circuit has \emph{polynomial depth} and the mapping \(\mathcal{T}\) is structured (e.g., low-rank/tensor-network/LoRA-style factors), QT can \emph{compress} a classical model with \(M\) trainable parameters to a quantum-generated parameterization whose trainable quantum degrees of freedom scale as
\[
K+\|\boldsymbol{\theta}\|\;=\;\mathcal{O}\!\big(\mathrm{poly}\!\left(\log M\right)\big),
\]
while maintaining competitive accuracy. Under mild smoothness/low-effective-rank assumptions on target weight tensors, there exist constructions where the QT parameterization \(\widehat{\mathbf{w}}\) approximates \(\mathbf{w}^\star\) with
\[
\|\widehat{\mathbf{W}}-\mathbf{W}^\star\|_F \le \varepsilon
\quad\text{using}\quad
K=\mathcal{O}\!\Big(\mathrm{poly}\!\big(\log M,\,\tfrac{1}{\varepsilon}\big)\Big),
\]
achieved by a polynomial-depth QNN and a tensorized \(\mathcal{T}\). In practice, structured \(\mathcal{T}\) (low-rank blocks, block-diagonal or convolutional kernels) bounds quantum-generated degrees of freedom while retaining expressivity; normalization/temperature scaling of \(\mathbf{g}\) and small-norm adapter constraints further stabilize training. Compared with direct QNN readouts, QT-NN preserves the efficiency and maturity of classical inference while injecting nonclassical correlations and offering an \(\mathcal{O}(\mathrm{poly}(\log M))\) trainable-parameter footprint on the quantum side.

\begin{table*}[!t]
\centering
\caption{Binary-IDS performance of all evaluated models. Bold indicates the best result for each metric.}
\label{tab:ids-results}
\small
\renewcommand{\arraystretch}{1.15}
\setlength{\tabcolsep}{4pt}             % minimal padding
\begin{tabularx}{\textwidth}{l C C C C C C C C}
\toprule
\textbf{Model} & \textbf{Qubits} & \textbf{Layers} &
\textbf{Class.\ Params} & \textbf{Quant.\ Params} &
\textbf{Accuracy} & \textbf{F1} & \textbf{Specificity} & \textbf{Sensitivity} \\
\midrule
SVM             & -  & -  & -    & -   & 0.924 & 0.951 & 0.836 & 0.947 \\
Quantum Kernel            & 8  & -  & -    & -   & 0.926 & 0.956 & 0.728 & 0.981 \\
QNN-6L          & 8  &  6 & 0    & 48  & 0.799 & 0.887 & 0.066 & \textbf{0.999} \\
QNN-8L          & 8  &  8 & 0    & 64  & 0.793 & 0.883 & 0.040 & 0.998 \\
QNN-10L         & 8  & 10 & 0    & 80  & 0.824 & 0.898 & 0.234 & 0.986 \\
HybridQNN-2L    & 8  &  2 & 18   & 16  & 0.926 & 0.952 & 0.896 & 0.934 \\
HybridQNN-4L    & 8  &  4 & 18   & 32  & 0.933 & 0.957 & 0.861 & 0.953 \\
HybridQNN-6L    & 8  &  6 & 18   & 48  & 0.930 & 0.954 & \textbf{0.904} & 0.937 \\
HybridQNN-8L    & 8  &  8 & 18   & 64  & \textbf{0.948} & \textbf{0.967} & 0.838 & 0.972 \\
HybridQNN-10L   & 8  & 10 & 18   & 80  & 0.944 & 0.965 & 0.920 & 0.942 \\
QT-NN (4, 2)    &  7 &  2 &   66* & 14  & 0.937 & 0.959 & 0.908 & 0.923 \\
QT-NN (8, 2)    &  8 &  2 &  162* & 16  & 0.939 & 0.960 & 0.808 & 0.940 \\
QT-NN (4, 4)    &  7 &  4 &   66* & 28  & 0.786 & 0.879 & 0.000 & \textbf{1.000} \\
QT-NN (8, 4)    &  8 &  4 &  162* & 32  & 0.938 & 0.959 & 0.881 & 0.925 \\
QT-NN (16, 4)   &  9 &  4 &  450* & 36  & 0.786 & 0.879 & 0.000 & \textbf{1.000} \\
\bottomrule
\end{tabularx}
\parbox{\textwidth}{\footnotesize
\textit{Note:} *: Classical parameters in QT-NN are not trained directly. They are generated by the quantum module controlled by the trainable quantum parameters.
}
\vskip -0.1in
\end{table*}

\section{Experiments}
\label{sec:experiments}
\subsection{Datasets and Feature Engineering}
We evaluate on UAVIDS-2025\cite{zenguavids2025}, a comprehensive benchmark for intrusion detection in UAV swarm networks. Traffic is generated in NS-3.24 with realistic UAV mobility via an extended BOID model and a Nakagami channel, using IEEE 802.11ac PHY/MAC and AODV routing. The corpus contains 122,171 labeled flow records across five categories: Normal, Blackhole, Flooding, Sybil, and Wormhole. Each flow is represented by 21 features grouped into (i) connection descriptors (e.g., duration, flags, hops), (ii) traffic-volume statistics (e.g., bytes/packets in/out, rates), and (iii) performance metrics (e.g., delay, jitter, retransmissions). The dataset explicitly supports evaluation under class imbalance and swarm mobility.

We distilled the original 22 raw flow attributes into an 8-dimensional feature vector that captures key temporal, structural, and statistical properties of each UAV network flow. Specifically: 

\begin{itemize}
    \item[(1)] \textbf{Flow duration:} \( T = t_{\text{last}} - t_{\text{first}} \), the elapsed time between the first and last packet of the flow.
    \item[(2)] \textbf{Packet rate:} \( r_p = N_{\text{pkt}} / T \), where \( N_{\text{pkt}} \) is the total number of packets transmitted in the flow.
    \item[(3)] \textbf{Byte rate:} \( r_b = B_{\text{tot}} / T \), where \( B_{\text{tot}} \) is the total number of bytes transmitted.
    \item[(4)] \textbf{Mean packet size:} \( \bar{s} = B_{\text{tot}} / N_{\text{pkt}} \), quantifying the average size of packets.
    \item[(5)] \textbf{Inter-arrival time dispersion:} \( \mathrm{CV}_{\Delta t} = \sigma_{\Delta t} / \mu_{\Delta t} \), the coefficient of variation of inter-packet times \( \Delta t_i = t_i - t_{i-1} \), measuring temporal variability.
    \item[(6)] \textbf{Directional asymmetry ratio (DAR):} 
    $\mathrm{DAR} = \bigl|B_{\rightarrow} - B_{\leftarrow}\bigr| \big/ \bigl(B_{\rightarrow} + B_{\leftarrow} + \epsilon\bigr)$,
    which quantifies traffic imbalance between forward and backward directions; \( \epsilon \) is a small constant to avoid division by zero.
    \item[(7)] \textbf{Short-term jitter:} 
    $J = \frac{1}{N_{\text{pkt}} - 2} \sum_{i} |\Delta t_{i+1} - \Delta t_{i}|$,
    
    capturing rapid changes in inter-arrival times that often signal congestion or attack activity.
    \item[(8)] \textbf{Peak-to-mean burst ratio (PMR):} 
    $\mathrm{PMR} = \max_{w} \bigl( r_b(w) \big/ r_b \bigr)$
    where \( r_b(w) \) denotes the local byte rate within sliding windows \( w \) (e.g., 100 ms), sensitive to short-duration traffic bursts.
\end{itemize}

To improve robustness and numerical stability, heavy-tailed rate-based features \( (r_p, r_b, \mathrm{PMR}) \) are log-transformed via \( \log(1+x) \). All features are then standardized to zero mean and unit variance using statistics computed solely from the training data to prevent data leakage. This compact and interpretable feature set retains sufficient discriminatory power to distinguish between normal, flooding (elevated \( r_p, r_b, \mathrm{PMR} \); low \( \mu_{\Delta t} \)), blackhole/wormhole (via directional imbalance in DAR and low \( r_b \)), and Sybil attacks (temporal irregularities captured by \( J \) and \( \mathrm{CV}_{\Delta t} \)), while ensuring compatibility with qubit-limited quantum machine learning models.

\subsection{QML Benchmark}

Classical SVM and its quantum-kernel analogue establish solid baselines with accuracy near 0.925 and F1-scores exceeding 0.95, yet their fixed kernels limit specificity under overlapping class distributions. Deep variational QNNs, although theoretically expressive, encounter barren-plateau trainability issues: as circuit depth increases, sensitivity approaches 0.99 while specificity collapses to as low as 0.04--0.23, resulting in overall accuracies approximately ten percentage points lower than classical models. Introducing lightweight classical post-processing---either via the parameter adapters in QT-NNs or the dual optimisation in QSVM---helps rebalance these trade-offs, yielding accuracies around 0.94, specificities above 0.80, and sensitivities above 0.90. The best performance is achieved by the eight-layer Hybrid QNN, which requires only eight qubits, 18 classical, and 64 quantum trainable parameters to reach an accuracy of 0.948, F1-score of 0.967, sensitivity of 0.972, and specificity of 0.838. This architecture leverages a shallow, hardware-efficient quantum encoder to capture high-order correlations in network flows, while a small classical head normalises and refines observables, mitigating gradient noise and over-activation. Compared to deeper QNNs, the Hybrid QNN avoids barren-plateau regions through reduced circuit width and depth; compared to kernel-based models, it retains task-specific learning capabilities. These results suggest that modest quantum expressivity, combined with lightweight classical regularisation, offers the most practical path to quantum advantage in UAV-swarm intrusion detection under current hardware constraints.

\section{Conclusion}
\label{sec:conclusion}
This study establishes a rigorous baseline for quantum intrusion detection in UAV swarms and clarifies when quantum resources translate into measurable benefit. Pure variational QNNs are currently hampered by trainability and noise, achieving extremely high sensitivities at the cost of false-positive inflation. Quantum-kernel SVMs reduce this variance but inherit the classical kernel’s expressivity ceiling. The hybrid architecture bridges the gap: by allocating correlation extraction to a shallow, hardware-efficient quantum circuit and calibration to a miniature classical head, the Hybrid QNN simultaneously maximises accuracy, F1, and resource efficiency, outperforming all other quantum and classical contenders under identical preprocessing and imbalance protocols. These findings point to hybrid-first design as the most practical path toward near-term quantum advantage in network-security workloads. Future work will extend the benchmark to federated and adversarially robust settings, incorporate real hardware runs, and explore automated circuit architecture search to further tighten the performance–resource trade-off.

\section*{Acknowledgement}

The author would like to thank Qingli Zeng for the helpful discussions regarding the dataset. This work was supported by the Engineering and Physical Sciences Research Council (EPSRC) under grant number EP/W032643/1.

\section*{Data Availability}
The UAVIDS-2025 dataset used in this study is publicly available as described in~\cite{zenguavids2025}. Tensor-network-based simulation results were produced using the cuTN-QSVM framework~\cite{chen2025validating}, which is available at \url{https://github.com/Tim-Li/cuTN-QSVM}.

\bibliographystyle{ieeetr}
\bibliography{references}

\begin{thebibliography}{10}

\bibitem{campion2018uav}
M.~Campion, P.~Ranganathan, and S.~Faruque, ``Uav swarm communication and control architectures: a review,'' {\em Journal of Unmanned Vehicle Systems}, vol.~7, no.~2, pp.~93--106, 2018.

\bibitem{zeng2018wireless}
T.~Zeng, M.~Mozaffari, O.~Semiari, W.~Saad, M.~Bennis, and M.~Debbah, ``Wireless communications and control for swarms of cellular-connected uavs,'' in {\em 2018 52nd Asilomar Conference on Signals, Systems, and Computers}, pp.~719--723, IEEE, 2018.

\bibitem{chen2020review}
X.~Chen, J.~Tang, and S.~Lao, ``Review of unmanned aerial vehicle swarm communication architectures and routing protocols,'' {\em Applied Sciences}, vol.~10, no.~10, p.~3661, 2020.

\bibitem{bertizzolo2020swarmcontrol}
L.~Bertizzolo, S.~D’oro, L.~Ferranti, L.~Bonati, E.~Demirors, Z.~Guan, T.~Melodia, and S.~Pudlewski, ``Swarmcontrol: An automated distributed control framework for self-optimizing drone networks,'' in {\em IEEE INFOCOM 2020-IEEE Conference on Computer Communications}, pp.~1768--1777, IEEE, 2020.

\bibitem{fotouhi2019survey}
A.~Fotouhi, H.~Qiang, M.~Ding, M.~Hassan, L.~G. Giordano, A.~Garcia-Rodriguez, and J.~Yuan, ``Survey on uav cellular communications: Practical aspects, standardization advancements, regulation, and security challenges,'' {\em IEEE Communications surveys \& tutorials}, vol.~21, no.~4, pp.~3417--3442, 2019.

\bibitem{menouar2017uav}
H.~Menouar, I.~Guvenc, K.~Akkaya, A.~S. Uluagac, A.~Kadri, and A.~Tuncer, ``Uav-enabled intelligent transportation systems for the smart city: Applications and challenges,'' {\em IEEE Communications Magazine}, vol.~55, no.~3, pp.~22--28, 2017.

\bibitem{bekmezci2013flying}
I.~Bekmezci, O.~K. Sahingoz, and {\c{S}}.~Temel, ``Flying ad-hoc networks (fanets): A survey,'' {\em Ad Hoc Networks}, vol.~11, no.~3, pp.~1254--1270, 2013.

\bibitem{asadpour2014micro}
M.~Asadpour, B.~Van~den Bergh, D.~Giustiniano, K.~A. Hummel, S.~Pollin, and B.~Plattner, ``Micro aerial vehicle networks: An experimental analysis of challenges and opportunities,'' {\em IEEE Communications Magazine}, vol.~52, no.~7, pp.~141--149, 2014.

\bibitem{motlagh2016low}
N.~H. Motlagh, T.~Taleb, and O.~Arouk, ``Low-altitude unmanned aerial vehicles-based internet of things services: Comprehensive survey and future perspectives,'' {\em IEEE internet of things journal}, vol.~3, no.~6, pp.~899--922, 2016.

\bibitem{choudhary2018intrusion}
G.~Choudhary, V.~Sharma, I.~You, K.~Yim, R.~Chen, and J.-H. Cho, ``Intrusion detection systems for networked unmanned aerial vehicles: A survey,'' in {\em 2018 14th International Wireless Communications \& Mobile Computing Conference (IWCMC)}, pp.~560--565, IEEE, 2018.

\bibitem{wang2024survey}
X.~Wang, Z.~Zhao, L.~Yi, Z.~Ning, L.~Guo, F.~R. Yu, and S.~Guo, ``A survey on security of uav swarm networks: attacks and countermeasures,'' {\em ACM Computing Surveys}, vol.~57, no.~3, pp.~1--37, 2024.

\bibitem{lopez2021towards}
M.~A. Lopez, M.~Baddeley, W.~T. Lunardi, A.~Pandey, and J.-P. Giacalone, ``Towards secure wireless mesh networks for uav swarm connectivity: Current threats, research, and opportunities,'' in {\em 2021 17th International Conference on Distributed Computing in Sensor Systems (DCOSS)}, pp.~319--326, IEEE, 2021.

\bibitem{santos2023flow}
L.~Santos, R.~Gon{\c{c}}alves, C.~Rabadao, and J.~Martins, ``A flow-based intrusion detection framework for internet of things networks,'' {\em Cluster Computing}, vol.~26, no.~1, pp.~37--57, 2023.

\bibitem{mrad2021federated}
I.~Mrad, L.~Samara, A.~A. Abdellatif, A.~Al-Abbasi, R.~Hamila, and A.~Erbad, ``Federated learning for uav swarms under class imbalance and power consumption constraints,'' in {\em 2021 IEEE Global Communications Conference (GLOBECOM)}, pp.~01--06, IEEE, 2021.

\bibitem{abbas2024challenges}
A.~Abbas, A.~Ambainis, B.~Augustino, A.~B{\"a}rtschi, H.~Buhrman, C.~Coffrin, G.~Cortiana, V.~Dunjko, D.~J. Egger, B.~G. Elmegreen, {\em et~al.}, ``Challenges and opportunities in quantum optimization,'' {\em Nature Reviews Physics}, pp.~1--18, 2024.

\bibitem{daley2022practical}
A.~J. Daley, I.~Bloch, C.~Kokail, S.~Flannigan, N.~Pearson, M.~Troyer, and P.~Zoller, ``Practical quantum advantage in quantum simulation,'' {\em Nature}, vol.~607, no.~7920, pp.~667--676, 2022.

\bibitem{cerezo2022challenges}
M.~Cerezo, G.~Verdon, H.-Y. Huang, L.~Cincio, and P.~J. Coles, ``Challenges and opportunities in quantum machine learning,'' {\em Nature computational science}, vol.~2, no.~9, pp.~567--576, 2022.

\bibitem{chen2025exploring}
L.-Y. Chen, T.-Y. Li, Y.-P. Li, N.-Y. Chen, and F.~You, ``Exploring chemical space with chemistry-inspired dynamic quantum circuits in the nisq era,'' {\em Journal of Chemical Theory and Computation}, 2025.

\bibitem{chen2025resource}
K.-C. Chen, F.~Burt, S.~Yu, C.-Y. Liu, M.-H. Hsieh, and K.~K. Leung, ``Resource-efficient compilation of distributed quantum circuits for solving large-scale wireless communication network problems,'' in {\em 2025 IEEE International Symposium on Circuits and Systems (ISCAS)}, pp.~1--5, IEEE, 2025.

\bibitem{chen2024quantum}
K.-C. Chen, X.~Xu, H.~Makhanov, H.-H. Chung, and C.-Y. Liu, ``Quantum-enhanced support vector machine for large-scale multi-class stellar classification,'' in {\em International Conference on Intelligent Computing}, pp.~155--168, Springer, 2024.

\bibitem{chen2020variational}
S.~Y.-C. Chen, C.-H.~H. Yang, J.~Qi, P.-Y. Chen, X.~Ma, and H.-S. Goan, ``Variational quantum circuits for deep reinforcement learning,'' {\em IEEE access}, vol.~8, pp.~141007--141024, 2020.

\bibitem{chen2022quantum}
S.~Y.-C. Chen, S.~Yoo, and Y.-L.~L. Fang, ``Quantum long short-term memory,'' in {\em ICASSP 2022-2022 IEEE International Conference on Acoustics, Speech and Signal Processing (ICASSP)}, pp.~8622--8626, IEEE, 2022.

\bibitem{chen2021federated}
S.~Y.-C. Chen and S.~Yoo, ``Federated quantum machine learning,'' {\em Entropy}, vol.~23, no.~4, p.~460, 2021.

\bibitem{chen2025compressedmediq}
K.-C. Chen, Y.-T. Li, T.-Y. Li, C.-Y. Liu, P.-H.~H. Lee, and C.-Y. Chen, ``Compressedmediq: Hybrid quantum machine learning pipeline for high-dimensional neuroimaging data,'' in {\em 2025 IEEE International Conference on Acoustics, Speech, and Signal Processing Workshops (ICASSPW)}, pp.~1--5, IEEE, 2025.

\bibitem{chen2025consensus}
K.-C. Chen, W.~Ma, and X.~Xu, ``Consensus-based distributed quantum kernel learning for speech recognition,'' in {\em 2025 IEEE International Conference on Acoustics, Speech, and Signal Processing Workshops (ICASSPW)}, pp.~1--5, IEEE, 2025.

\bibitem{chen2024quantum2}
K.-C. Chen, X.~Li, X.~Xu, Y.-Y. Wang, and C.-Y. Liu, ``Quantum-classical-quantum workflow in quantum-hpc middleware with gpu acceleration,'' in {\em 2024 International Conference on Quantum Communications, Networking, and Computing (QCNC)}, pp.~304--311, IEEE, 2024.

\bibitem{chen2025validating}
K.-C. Chen, T.-Y. Li, Y.-Y. Wang, S.~See, C.-C. Wang, R.~Wille, N.-Y. Chen, A.-C. Yang, and C.-Y. Lin, ``Validating large-scale quantum machine learning: Efficient simulation of quantum support vector machines using tensor networks,'' {\em Machine Learning: Science and Technology}, vol.~6, no.~1, p.~015047, 2025.

\bibitem{liu2024towards}
J.~Liu, M.~Liu, J.-P. Liu, Z.~Ye, Y.~Wang, Y.~Alexeev, J.~Eisert, and L.~Jiang, ``Towards provably efficient quantum algorithms for large-scale machine-learning models,'' {\em Nature Communications}, vol.~15, no.~1, p.~434, 2024.

\bibitem{ma2025robust}
W.~Ma, K.-C. Chen, S.~Yu, M.~Liu, and R.~Deng, ``Robust decentralized quantum kernel learning for noisy and adversarial environment,'' {\em arXiv preprint arXiv:2504.13782}, 2025.

\bibitem{chen2025noise}
K.-C. Chen, M.~Prest, F.~Burt, S.~Yu, and K.~K. Leung, ``Noise-aware detectable byzantine agreement for consensus-based distributed quantum computing,'' in {\em 2025 International Conference on Quantum Communications, Networking, and Computing (QCNC)}, pp.~210--215, IEEE, 2025.

\bibitem{hsu2025quantum}
Y.-C. Hsu, T.-Y. Li, and K.-C. Chen, ``Quantum kernel-based long short-term memory,'' in {\em 2025 IEEE International Conference on Acoustics, Speech, and Signal Processing Workshops (ICASSPW)}, pp.~1--5, IEEE, 2025.

\bibitem{abbas2021power}
A.~Abbas, D.~Sutter, C.~Zoufal, A.~Lucchi, A.~Figalli, and S.~Woerner, ``The power of quantum neural networks,'' {\em Nature Computational Science}, vol.~1, no.~6, pp.~403--409, 2021.

\bibitem{yu2024shedding}
S.~Yu, Z.~Jia, A.~Zhang, E.~Mer, Z.~Li, V.~Crescimanna, K.-C. Chen, R.~B. Patel, I.~A. Walmsley, and D.~Kaszlikowski, ``Shedding light on the future: Exploring quantum neural networks through optics,'' {\em Advanced Quantum Technologies}, p.~2400074, 2024.

\bibitem{liu2024quantum}
C.-Y. Liu, E.-J. Kuo, C.-H.~A. Lin, J.~G. Young, Y.-J. Chang, M.-H. Hsieh, and H.-S. Goan, ``Quantum-train: Rethinking hybrid quantum-classical machine learning in the model compression perspective,'' {\em Quantum Machine Intelligence}, vol.~7, no.~2, p.~80, 2025.

\bibitem{liu2025quantum}
C.-Y. Liu, C.-H.~A. Lin, and K.-C. Chen, ``Quantum-train with tensor network mapping model and distributed circuit ansatz,'' in {\em ICASSP 2025-2025 IEEE International Conference on Acoustics, Speech and Signal Processing (ICASSP)}, pp.~1--4, IEEE, 2025.

\bibitem{liu2025programming}
C.-Y. Liu, S.~Y.-C. Chen, K.-C. Chen, W.-J. Huang, and Y.-J. Chang, ``Programming variational quantum circuits with quantum-train agent,'' in {\em 2025 International Conference on Quantum Communications, Networking, and Computing (QCNC)}, pp.~544--548, IEEE, 2025.

\bibitem{chen2025quantum}
K.-C. Chen, S.~Y.-C. Chen, C.-Y. Liu, and K.~K. Leung, ``Quantum-train-based distributed multi-agent reinforcement learning,'' in {\em 2025 IEEE Symposium for Multidisciplinary Computational Intelligence Incubators (MCII Companion)}, pp.~1--5, IEEE, 2025.

\bibitem{zenguavids2025}
Q.~Zeng, A.~Bashir, and F.~Nait-Abdesselam, ``Uavids-2025: A benchmark dataset for intrusion detection in uav networks using machine learning techniques,'' in {\em 2025 IEEE Conference on Communications and Network Security (CNS)}, IEEE, 2025.

\end{thebibliography}

\end{document}